\documentclass[prd,twocolumn,superscriptaddress,showpacs,preprintnumbers,bibnotes,nofootinbib]{revtex4}
\usepackage{amsmath,,amssymb,amsfonts,graphicx,epsfig}

\RequirePackage{slashed} 

\def\a{\alpha}
\def\d{\delta}
\def\g{\gamma}
\def\ve{\varepsilon}

\def\D{\Delta}
\def\G{\Gamma}
\def\L{\Lambda}
\def\hs{\hspace}
\def\ol{\overline}
\def\no{\nonumber}
\def\ua{\uparrow}
\def\da{\downarrow}
\def\lra{\longrightarrow}
\def\lf{\left}
\def\rg{\right}
\def\la{\langle}
\def\ra{\rangle}

\begin{document}

\preprint{PHY-11852-TH-2007}

\title{Transversity quark distributions in a covariant quark-diquark model}

\author{I.~C.~Clo\"et}
\email{icloet@anl.gov}
\affiliation{Physics Division, Argonne National Laboratory, Argonne, IL 60439-4843, U.S.A.}
\author{W.~Bentz}
\email{bentz@keyaki.cc.u-tokai.ac.jp}
\affiliation{Department of Physics, School of Science, Tokai University, 
                         Hiratsuka-shi, Kanagawa 259-1292, Japan}
\author{A.~W.~Thomas}
\email{awthomas@jlab.org}
\affiliation{Jefferson Lab, 12000 Jefferson Avenue, Newport News, VA 23606, U.S.A.}
\affiliation{College of William and Mary, Williamsburg, VA 23187, U.S.A.}

\begin{abstract}
Transversity quark light-cone momentum distributions are calculated for the nucleon. 
We utilize a modified Nambu--Jona-Lasinio model in which confinement is simulated 
by eliminating unphysical thresholds for nucleon decay into quarks. The nucleon bound state
is obtained by solving the relativistic Faddeev equation in the quark-diquark approximation, where both 
scalar and axial-vector diquark channels are included. Particular attention is paid to 
comparing our results with
the recent experimental extraction of the transversity distributions by Anselmino \textit{et al}.
We also compare our transversity results with earlier spin-independent and helicity
quark distributions calculated in the same approach.
\end{abstract}

\maketitle

\section{Introduction}

For perfectly collinear quarks, at leading twist, the nucleon has three 
independent quark distribution functions for each quark flavour: the unpolarized 
distributions, $q(x)$, the helicity distributions, $\D q(x)$, and the transversity 
distributions, $\D_T q(x)$. Knowledge of each distribution is of equal importance
if one is to have a robust description of nucleon structure.
The unpolarized and helicity distributions have been extensively studied, both
experimentally and theoretically, for many years \cite{Buras:1979yt,Anselmino:1994gn}.
However interest in the transversity distributions is relatively recent 
\cite{Barone:2001sp}.

The transversity distributions are associated with quark-nucleon forward
Compton amplitudes, where both the quark and nucleon helicities are flipped.
Hence these distributions are chiral-odd. The electroweak 
and strong interactions conserve chirality, so the transversity distributions must couple
to another chiral-odd quantity in scattering cross-sections. 
This is not possible in inclusive deep inelastic scattering (DIS) \cite{Barone:2001sp}. 
However, for certain semi-inclusive DIS 
\cite{Collins:1992kk,Jaffe:1996wp,Barone:2001sp}
and Drell-Yan processes \cite{Ralston:1979ys,Artru:1989zv,Jaffe:1991ra,Barone:2001sp}
the transversity distributions do appear at leading twist 
in the cross-section. For semi-inclusive DIS the transverse quark distributions couple 
to particular chiral-odd fragmentation functions (Collins functions), whereas in Drell-Yan 
dilepton production they appear with either a transverse quark or anti-quark
distribution from the partner hadron. 

The transversity distributions are particularly interesting for several reasons,
for example: 
\begin{itemize}
\item the moments of the transversity valence distributions are related to the nucleon
tensor charge \cite{Jaffe:1991kp};
\item the quark bilinears associated with transversity are
odd under charge conjugation and hence the quark--anti-quark sea does not contribute
\cite{Jaffe:1991kp};
\item helicity conservation at the quark-gluon vertex prevents 
mixing between the quark and gluon transversity distributions under QCD evolution 
\cite{Artru:1989zv,Hayashigaki:1997dn,Vogelsang:1997ak};
\item gluon transversity distributions, $\D_T g(x)$, only exist for targets
with $J \geqslant 1$, because measurement of a gluon transversity
distribution requires that the target change helicity by two units
of angular momentum and this is not possible for spin-$\tfrac{1}{2}$ targets
\cite{Collins:1992kk,Artru:1989zv,Barone:2001sp}. 
\end{itemize}
These results imply that the transversity distributions are valence quark dominated and 
evolve as non-singlets under DGLAP evolution, where the angular momentum generated
by the DGLAP kernels is not shared between the quark and gluon sectors. 
These features of the transversity distributions make them particularly amenable to 
a quark model treatment. 

In this paper we calculate the transversity distributions for the proton
using a Nambu--Jona-Lasinio model \cite{Nambu:1961tp,Nambu:1961fr}. This model is 
attractive because it is covariant and has a transparent description of spontaneous chiral
symmetry breaking. Confinement -- in the sense that there exists no threshold for
nucleon decay into quarks -- is also implemented via the regularization
procedure \cite{Ebert:1996vx,Hellstern:1997nv}. We construct the nucleon as a bound 
state solution of the relativistic Faddeev equation \cite{Ishii:1993xm}, in the 
quark-diquark approximation \cite{Asami:1995xq,Buck:1992wz}, where both scalar and 
axial-vector diquark channels are included. We compare our transversity results to
spin-independent and helicity quark distributions calculated in the same approach \cite{Cloet:2005pp}. 
Particular attention is also paid to a comparison of our transversity results with the recent, and to 
date the only, experimental extraction of the transversity distributions by 
Anselmino \textit{et al.}, presented in Ref.~\cite{Anselmino:2007fs}.

\section{Transversity Quark distributions}

The transversity distributions render a probability interpretation 
analogous to the other two leading twist distributions. In a transversely polarized hadron
they represent the number density of quarks in an eigenstate of the transverse
Pauli-Lubanski operator, $\slashed{S}_\perp\g_5$, with eigenvalue $+\tfrac{1}{2}$,
minus the number density of quarks with eigenvalue $-\tfrac{1}{2}~$, \cite{Jaffe:1991ra}
that is
\begin{equation}
\D_T q(x) = q_\ua(x) - q_\da(x).
\end{equation}
In an helicity basis the helicity distributions are expressed as
\begin{equation}
\D q(x) = q_+(x) - q_-(x),
\end{equation}
where $q_+(x)$ is the number density of quarks with helicity parallel to the 
hadron helicity and $q_-(x)$ is the quark number density with helicity 
anti-aligned. The spin-independent distributions in each basis are given 
by
\begin{equation}
q(x) = q_\ua(x) + q_\da(x) = q_+(x) + q_-(x).
\label{eq:spinIN}
\end{equation}

The leading twist quark distribution functions are defined by light-cone Fourier 
transforms of connected matrix elements of particular quark field bilinears. 
For example, the twist-2 transversity distribution is defined by
\begin{multline}
\D_T q(x) = p^+\int \frac{d\xi^-}{2\pi} e^{i\,x\,p^+\xi^-} \\
\la p,s\lvert \overline{\psi}_q(0)\g^+\g^1\g_5\psi_q(\xi^-) \rvert p,s\ra_c,
\label{eq:trans}
\end{multline}
where $\psi_q$ is a quark field of flavour $q$ and $x$ is the Bjorken scaling
variable.\footnote{The formal expressions for the spin-independent and helicity
distributions can be obtained from Eq.~\eqref{eq:trans} via the operator 
replacements $\g^+\g^1\g_5 \to \g^+$ and $\g^+\g^1\g_5 \to \g^+\g_5$, respectively.
The nucleon state is normalized according to the non-covariant
light-cone normalization, namely
$\la p,s\lvert \overline{\psi}_u\g^+ \psi_u + \overline{\psi}_d\g^+ \psi_d\rvert p,s\ra_c =3$.}
In Eq.~\eqref{eq:trans} the target polarization is in the $x$-direction,
with the $z$-direction defined by the photon 3-momentum. 

\begin{figure}[tbp]
\centering\includegraphics[height=2.10cm,angle=0]{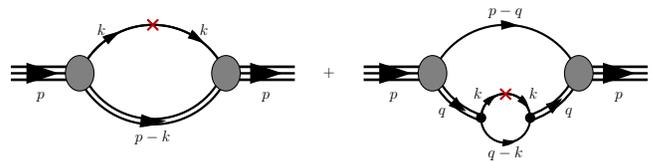}
\caption{Feynman diagrams representing the transversity quark distributions in the 
nucleon, needed in the evaluation of Eq.~(\ref{eq:transTr}). The single line
represents the quark propagator and the double line the 
diquark $t$-matrix. The shaded oval denotes the quark-diquark vertex function and the 
operator insertion has the form 
$\g^+\g^1\g_5\,\d\!\!\lf(x - \frac{k^+}{p^+}\rg)\frac{1}{2}\lf(1\pm\tau_z\rg)$.}
\label{fig:feydiagrams}
\end{figure}

The evaluation of the quark distributions is facilitated by expressing 
Eq.~\eqref{eq:trans} in the form \cite{Jaffe:1983hp,Jaffe:1985je}
\begin{equation}
 \D_T q(x) = -i \int\frac{d^4k}{(2\pi)^4} \d\lf(x-\frac{k^+}{p^+}\rg)
\textrm{Tr}\lf[\g^+\g^1\g_5\,M(p,k)\rg],
\label{eq:transTr}
\end{equation}
where $M(p,k)$ is the quark two-point function in a nucleon.
The quark distributions can then be expressed in terms of Feynman diagrams
for any model where the nucleon is represented by a bound state of quarks. The diagrams 
we consider are given in Fig.~\ref{fig:feydiagrams}. In our pure valence quark model
there should also be a third diagram, the so-called quark exchange term \cite{Asami:1995xq}, 
however this diagram does not contribute within the static approximation used here
\cite{Buck:1992wz,Ishii:1993xm}. 

In the Feynman diagrams of Fig.~\ref{fig:feydiagrams} the single line represents 
a constituent quark propagator and the double line a diquark $t$-matrix.
The diagram on the left is referred to as the quark diagram and on
the right we have the diquark diagram, where we include both scalar and axial-vector 
diquarks. Separating the isospin coefficients, the 
$u$- and $d$-quark transversity distributions can be expressed as
\begin{align}
\label{eq:delu}
\D_T u_v(x) &= \D_T f^s_{q/N}(x)  \no \\
&\hs{-7mm} + \frac{1}{2}\,\D_T f^s_{q(D)/N}(x) + \frac{1}{3}\, \D_T f^a_{q/N}(x) \no \\
&\hs{-1mm}+ \frac{5}{6}\,\D_T f^a_{q(D)/N}(x) + \frac{1}{2\sqrt{3}} \D_T f_{q(D)/N}^m (x),  \\ 
\label{eq:deld}
\D_T d_v(x) &= \frac{1}{2}\,\D_T f^s_{q(D)/N}(x) + \frac{2}{3}\,\D_T f^a_{q/N}(x)  \no \\
&\hs{0mm} + \frac{1}{6}\,\D_T f^a_{q(D)/N}(x) - \frac{1}{2\sqrt{3}} \D_T f_{q(D)/N}^m (x),
\end{align}
where each term represents a particular Feynman diagram in Fig.~\ref{fig:feydiagrams}.
The superscripts $s$, $a$ and $m$ refer to the scalar, axial-vector or mixing
terms, respectively. The subscript $q/N$ implies a quark diagram and
$q(D)/N$ a diquark diagram. Because the scalar diquark has spin zero, 
we have $\D_T f^s_{q(D)/N}(x)=0$ and hence the
transversity of the $d$-quark arises exclusively from 
the axial-vector and the mixing terms. 

Importantly, in this covariant framework, the Ward identities corresponding to
number and momentum conservation are satisfied, guaranteeing the validity of the  
baryon number and momentum sum rules \cite{Mineo:1999eq,Mineo:2003vc} from the 
outset.

\section{The nucleon in the NJL model}

The NJL model which we employ to determine the transversity distributions is
discussed in Ref.~\cite{Cloet:2005pp}, where it was used to determine the spin-independent
and helicity quark distribution functions. Therefore we 
give only a brief outline of the model in this section and refer the reader to
Ref.~\cite{Cloet:2005pp} for further details.

We represent the nucleon as a quark--diquark bound state,
where both the scalar and axial-vector diquark channels are included. 
The nucleon $t$-matrix is obtained as the solution of the relativistic Faddeev equation 
\begin{equation}
T =  Z + Z\,\Pi_N\,T =  Z + T\,\Pi_N\,Z.
\label{eq:t}
\end{equation}
In the static approximation the quark exchange kernel, $Z$,  
becomes \cite{Ishii:1993xm}
\begin{equation}
Z = \frac{3}{M} \begin{pmatrix} 
                   1                 & \sqrt{3}\g^\mu\,\g_5 \\
                 \sqrt{3}\g_5\,\g^\mu & -\g^\mu\,\g^{\mu'} \\
                \end{pmatrix},
\end{equation}
where $M$ is the constituent quark mass.
With this approximation, $\Pi_N$ of Eq.~\eqref{eq:t} effectively becomes the quark-diquark
bubble graph, given by
\begin{align}
\Pi_N(p) &= \int \frac{d^4k}{(2\pi)^4}\, \tau(p-k)\, S(k),
\label{eq:nucleonbubble}
\end{align}
where
\begin{align}
\tau(q) =  \begin{pmatrix} \tau_s(q) & 0 \\ 0 & \tau_a^{\mu\nu}(q) \end{pmatrix}.
\label{tau}
\end{align}
The quantities $\tau_s(q)$ and $\tau_a^{\mu\nu}(q)$ are the solutions to 
the NJL scalar and axial-vector diquark Bethe-Salpeter equations, respectively 
\cite{Ishii:1993xm,Ishii:1995bu,Mineo:2002bg}, and $S(k)$ in Eq.~\eqref{eq:nucleonbubble}
is the usual fermion propagator for a constituent quark.

In the non-covariant light-cone normalization used already
in Eq.~(\ref{eq:trans}), the quark-diquark vertex function, $\G_N$, is defined by the 
behaviour of $T$ near the pole
\begin{align} 
T \stackrel{p_+ \to \ve_p}{\lra} \frac{\G_N\,\ol{\G}_N}{p_+ - \ve_p},
\label{pole}
\end{align}
where $\varepsilon_p = \frac{M_N^2}{2p_-}$ is the nucleon's light-cone energy. 
The vertex functions satisfy homogeneous Faddeev equations of the 
form
\begin{align}
\label{eq:f}
\G_N = Z\,\Pi_N \, \G_N, \quad \text{and} \quad
\ol{\G}_N = \ol{\G}_N\, \Pi_N \,Z.
\end{align}

As the NJL model is non-renormalizable, a regularization prescription
must be specified to fully define the theory.
We choose the proper-time regularization
scheme \cite{Schwinger:1951nm,Ebert:1996vx,Hellstern:1997nv,Bentz:2001vc,Bijnens:1992uz}, 
where loop integrals of products of propagators are evaluated by
introducing Feynman parameters, Wick rotating and making the denominator 
replacement 
\begin{equation}
\frac{1}{X^n} 
\lra
\frac{1}{(n-1)!}\,\int_{1/(\L_{UV})^2}^{1/(\L_{IR})^2}\,d\tau\,\tau^{n-1}\,e^{-\tau\,X},
\end{equation}
where $\L_{IR}$ and $\L_{UV}$ are, respectively, the infrared  and ultraviolet cutoffs.
The former has the effect of eliminating unphysical thresholds for hadron decay into
quarks, hence simulating an important aspect of quark confinement \cite{Hellstern:1997nv}.
The proper-time scheme also preserves the gauge invariance and 
Poincar\'e covariance of the theory.

\section{Results}

The six parameters of the model are: the regularization parameters,
$\L_{IR}$ and $\L_{UV}$, the current quark mass, $m$, and three quark-quark coupling
constants $G_\pi$, $G_s$ and $G_a$ \cite{Cloet:2005pp}. 
The infrared or confinement scale is expected to be of 
order $\L_\text{QCD}$ and we choose $\L_{IR} = 240\,$MeV, however our results 
exhibit only a minor dependence on $\L_{IR}$ when it is varied between 200$-$300$\,$MeV.
The parameters $m$, $\L_{UV}$ and the $q\bar{q}$ coupling in the pion channel, $G_\pi$, 
are determined by requiring 
$M=400\,$MeV via the gap equation, $f_\pi=93\,$MeV from the
one loop pion decay diagram and $m_\pi=140\,$MeV from 
the pole of the $q \overline{q}$ $t$-matrix
in the pion channel. This gives $m=16.4\,$MeV, 
$\L_{UV} = 645\,$MeV and $G_\pi = 19.04\,$GeV$^{-2}$.
The $qq$ couplings in the scalar, $G_s$, and axial-vector, $G_a$, diquark 
channels are determined by reproducing the nucleon mass,
$M_N = 940\,$MeV, as the solution of Eq.~\eqref{eq:f} and 
satisfying the Bjorken sum rule within our model, with $g_A = 1.267$. 
We obtain $G_s = 7.49~$GeV$^{-2}$ and $G_a = 2.80~$GeV$^{-2}$. 
With these model parameters the diquark masses 
are $M_s= 687\,$MeV and $M_a=1027\,$MeV.

Inherent in most model determinations of the quark distributions is the absence
of an explicit $Q^2$ scale. Consequently the model
scale, $Q_0^2$, must be determined via comparison with empirical results.
We do this by optimizing $Q_0^2$ such that the spin-independent 
valence up quark distribution, $u_v(x) \equiv u(x) - \bar{u}(x)$, 
best reproduces the empirical parametrization after NLO $Q^2$ 
evolution.\footnote{The DGLAP evolution of our results
for the three twist-2 quark distributions is performed using the appropriate computer 
program from Refs.~\cite{Miyama:1995bd,Hirai:1997gb,Hirai:1997mm}.
For the DGLAP parameters we choose $N_f=3$ and $\Lambda_{\text{QCD}}=250~$MeV.}
We find $Q_0^2 = 0.16~$GeV$^2$ \cite{Cloet:2005pp}, which is typical of
valence dominated models \cite{Mineo:2002bg,Mineo:1999eq,Schreiber:1991tc}.

\begin{figure}[tbp]
\centering\includegraphics[width=\columnwidth,clip=true,angle=0]{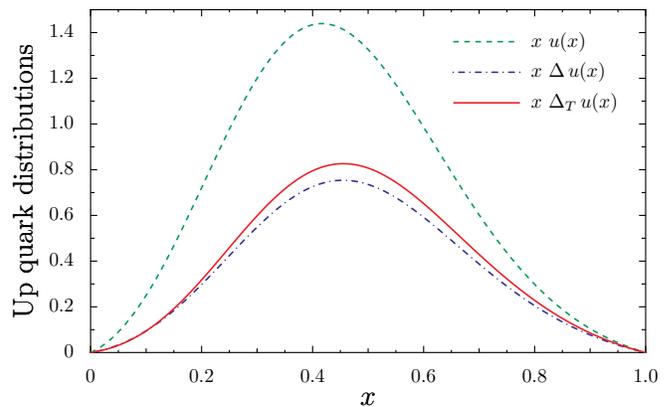}
\caption{Model results for the triplet of twist-2 valence up quark distributions,
at $Q_0^2 = 0.16\,$GeV$^2$.
The spin-independent and helicity distributions are taken from earlier work
presented in Ref.~\cite{Cloet:2005pp}.}
\label{fig:freeup}
\end{figure}

\begin{figure}[tbp]
\centering\includegraphics[width=\columnwidth,clip=true,angle=0]{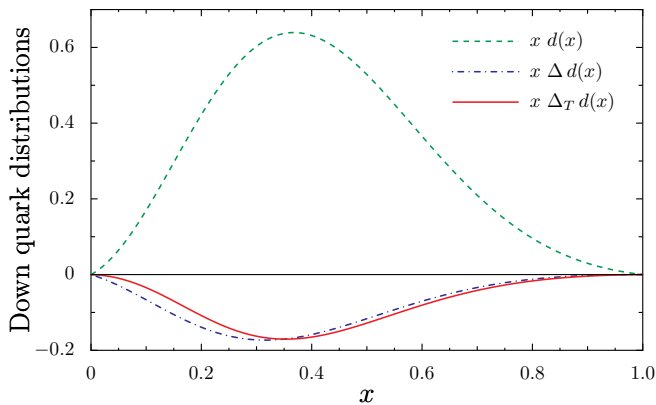}
\caption{Model results for the triplet of twist-2 valence down quark distributions,
at $Q_0^2 = 0.16\,$GeV$^2$. The spin-independent and helicity results are taken 
from earlier work presented in Ref.~\cite{Cloet:2005pp}.}
\label{fig:freedown}
\end{figure}

In Figs.~\ref{fig:freeup} and \ref{fig:freedown} we illustrate our results for the transversity 
valence $u$- and $d$-quark distributions, respectively. Included in these figures are
our results for the spin-independent and helicity valence distributions taken
from Ref.~\cite{Cloet:2005pp}. We find a transversity $u$-quark distribution
slightly larger than the corresponding helicity distribution for all $x$. The transversity
and helicity $d$-quark distributions are very similar for $x \gtrsim 0.35$, however
at smaller $x$ the transversity distribution is noticeably suppressed. The difference
between the transversity and helicity valence distributions results purely from relativistic
effects, since in the non-relativistic limit, where boosts and rotations commute, there can be 
no preferential polarization direction and therefore the helicity and transversity distributions 
must be equal. This is true only at the model scale as the helicity and transversity distributions
evolve differently under DGLAP evolution.

\begin{figure}[tbp]
\centering\includegraphics[width=\columnwidth,clip=true,angle=0]{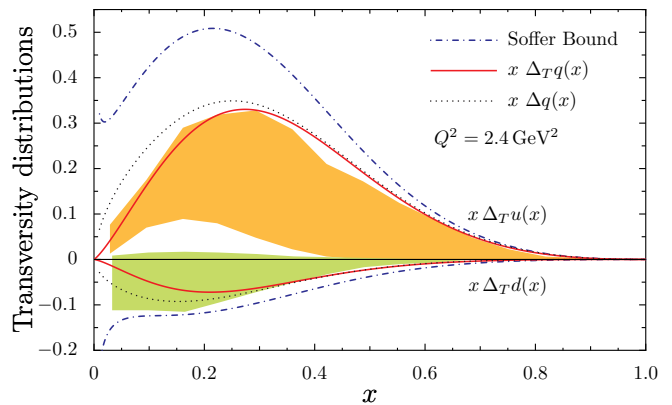}
\caption{The shaded areas are the empirical results of Ref.~\cite{Anselmino:2007fs}, with a one-sigma
confidence interval.
The dot-dashed line is the GRV Soffer bound \cite{Gluck:1998xa,Gluck:2000dy}
and the solid lines are our results for the transversity distributions. 
The dotted curves are our helicity distributions taken from Ref.~\cite{Cloet:2005pp}.
All results are at $Q^2 = 2.4\,$GeV$^2$.}
\label{fig:transexper}
\end{figure}

The first experimental extraction of the transversity distributions was achieved
only recently, and is published in Ref.~\cite{Anselmino:2007fs}. The authors combined
semi-inclusive DIS data from HERMES and COMPASS with $e^+e^-$ annihilation data 
from BELLE to simultaneously extract the transversity distributions and Collins 
functions. Their results for the transversity $u$- and $d$-quark distributions
at $Q^2 = 2.4\,$GeV$^2$ are presented in Fig.~\ref{fig:transexper} as the shaded 
regions, which represent a one-sigma confidence interval. Included in this
figure are our valence results at $Q^2 = 2.4\,$GeV$^2$ and the empirical Soffer bound 
at the same scale obtained from an evolution of the GRV parametrizations given in
Refs.~\cite{Gluck:1998xa,Gluck:2000dy}. A direct comparison between our results
and those of Ref.~\cite{Anselmino:2007fs} should be valid for $x > 0.2$ where
transversity anti-quark distributions are expected to be small.
We find that our results lie slightly
outside the one-sigma error bounds of the empirical parametrizations for 
$x \gtrsim 0.3$. On the same figure we illustrate our results for the helicity
distributions at $Q^2 = 2.4\,$GeV$^2$. At this scale our helicity and transversity
distributions remain very similar for $x \gtrsim 0.4$, however the differing
$Q^2$ evolution has resulted in a substantial suppression for the transversity distributions
at smaller $x$, when compared with the results at the model scale. The helicity 
distributions given in Fig.~\ref{fig:transexper} are in excellent agreement with the empirical 
parametrizations \cite{Cloet:2005pp}.

The first moments of the transversity valence distributions are related to the nucleon's
isovector tensor charge, $g_T$, via \cite{Jaffe:1991ra}
\begin{equation}
\int_0^1 dx \lf[\D_T u_v(x,Q^2) - \D_T d_v(x,Q^2)\rg] = g_T(Q^2).
\label{eq:tensor}
\end{equation}
The nucleon's isoscalar tensor charge,
$g_T^0$, is defined as the sum of the valence transversity moments.
For these moments we obtain $\D_T u_v = 1.04$ and $\D_T d_v = -0.24$, giving
a nucleon isovector tensor charge of $g_T = 1.28$ and a isoscalar charge of $g_T^0 = 0.80$,
at $Q^2 = 0.16\,$GeV$^2$. At the GRV scale of $Q^2 = 0.4\,$GeV$^2$ we 
obtain $\D_T u_v = 0.69$, $\D_T d_v = -0.16$, $g_T = 0.85$ and $g_T^0 = 0.53$.\footnote{These
values are obtained using the NLO result \cite{Barone:2001sp}
\begin{multline}
\D_T q(Q^2) = \lf[\frac{\a_s(Q^2)}{\a_s(Q^2_0)}\rg]^{\frac{4}{27}}\\
\lf[1-\frac{337}{486\pi}\lf[\a_s(Q^2_0)-\a_s(Q^2)\rg]\rg]\, \D_T q(Q^2_0).
\label{eq:evoltran}
\end{multline}
}
It is well known that the first moments of the helicity distributions
give the quark spin content of the nucleon.
However, the first moments of the transversity distributions
are not equivalent to the quark spin in the transverse direction, since the
expectation value of $\g^1\g_5$ is not equal to that of $\Sigma_\perp=\g^0\g^1\g_5$.
That is, the isoscalar tensor charge cannot be interpreted as a transverse 
spin sum \cite{Jaffe:1991ra,Jaffe:1991kp}.

The paper of Anselmino \textit{et al.}\ does not give explicit values for the
transversity moments, however if we integrate their parametrizations and accurately 
determine the errors from Fig. 7 in Ref.~\cite{Anselmino:2007fs}, we find
$\D_T u = 0.46^{+0.36}_{-0.28}$ and  $\D_T d = -0.19^{+0.30}_{-0.23}$ at $Q^2 = 0.4\,$GeV$^2$.\footnote{These 
values are obtained from the FIT-I results given in Table~I of Ref.~\cite{Anselmino:2007fs}. 
The FIT-II results give even smaller moments, namely $\D_T u = 0.40$ and  $\D_T d = -0.16$. The errors
quoted in the text are first obtained at $Q^2 = 2.4\,$GeV$^2$, then the NLO result expressed in
Eq.~\eqref{eq:evoltran}
is used to devolve to the parametrization scale, $Q^2 = 0.4\,$GeV$^2$. We find that
this procedure works very well for the central value, which can be determined at both $Q^2$ scales.} 
Using these moments their tensor charges are $g_T \sim 0.65$ and $g_T^0 \sim 0.27$, 
assuming $\D_T\bar{u}-\D_T\bar{d} \simeq 0$.

The similarity between the transversity and helicity distributions, which we illustrate
in Fig.~\ref{fig:transexper},  is a common theme for model calculations --
see Refs.~\cite{Jaffe:1991ra,Scopetta:1997qg,Barone:1996un,Gamberg:1998vg,Suzuki:1997vu,
Suzuki:1997wv,Ma:1997gy}. 
This should come as no surprise since the constituent quark masses are typically rather large,
$350-450\,$MeV, in these calculations and therefore enormous relativistic effects 
would be unexpected. For the moment, we have no more than a hint that this may
disagree with the experimental extraction of 
Anselmino \textit{et al.}, who find a $u$-quark transversity distribution that may in fact be
far smaller than its helicity counterpart. This difference is emphasized if we study the
moments. The empirical values for the helicity moments are $\D u = 0.851 \pm 0.075$ and $\D d = -0.415\pm 0.124$ 
\cite{Bluemlein:2002be}, for the valence distributions the results are
$\D u_v = 0.926 \pm 0.071$ and $\D d_v = -0.341\pm 0.123$ \cite{Bluemlein:2002be}.
Therefore $\D_T u$ extracted from the fit of Ref.~\cite{Anselmino:2007fs}
is potentially far smaller than the $u$-quark helicity moment.
If this difference survives it would indicate a difficulty for relativistic model 
approaches based on the concept of a constituent quark mass.

Within our NJL model calculation we find $\D u_v = 0.97$ and $\D d_v = -0.30$,\footnote{Note, 
the helicity moments given  here differ slightly from those in Ref.~\cite{Cloet:2005pp} 
as in this work the infrared cutoff is slightly different.} which differ only slightly
from the transversity results given earlier. 
The nine models discussed in the transversity review by Barone \textit{et al.}\ 
\cite{Barone:2001sp} all find $\D_T u > \D u$ and $\D_T d \sim \D d$ at the model scale.
This consensus amongst the various model approaches, together with our results,
contrasted with the potentially small empirical $u$-quark transversity parametrization given in 
Ref.~\cite{Anselmino:2007fs}, presents an interesting problem. We therefore eagerly
await new experiment data so that the empirical transversity distributions can be 
better constrained.

\begin{figure}[tbp]
\centering\includegraphics[width=\columnwidth,clip=true,angle=0]{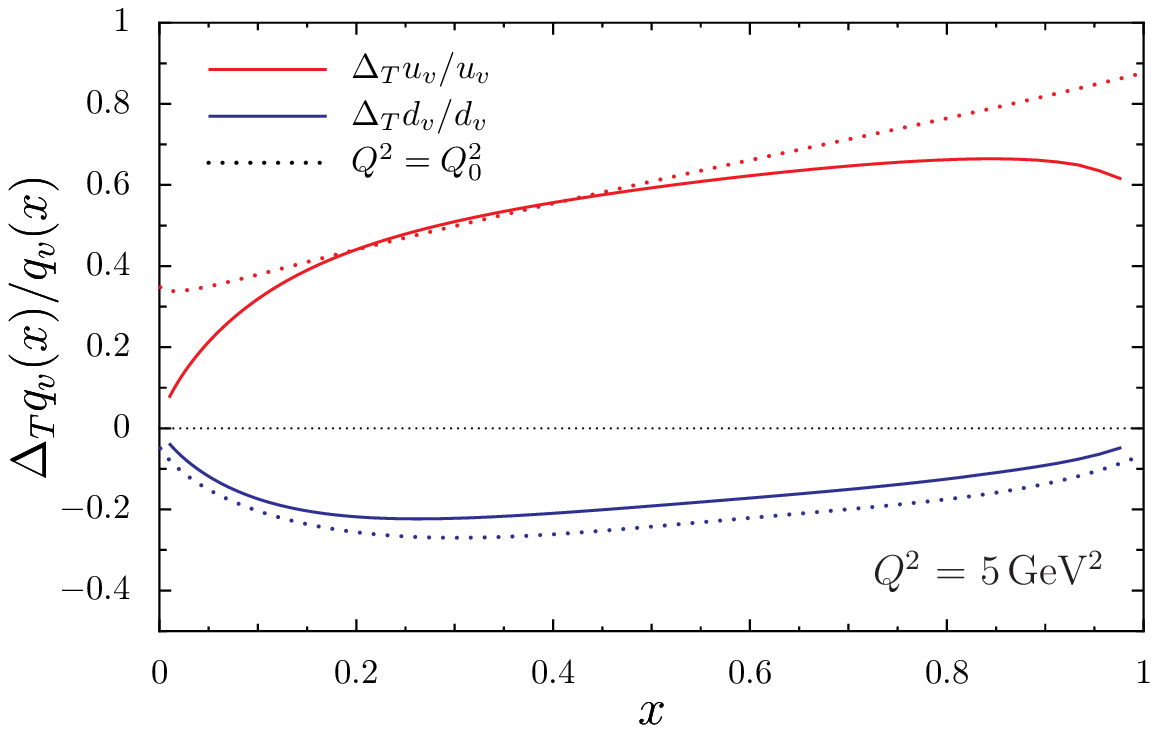}
\caption{Quark distribution ratios, in each case the dotted line is the result 
before QCD evolution and the solid line is the result for $Q^2=5\,$GeV$^2$.}
\label{fig:ratio2}
\end{figure}

The Soffer inequality  \cite{Soffer:1994ww}, $q(x) + \D q(x) \geqslant 2 \lf|\D_T q(x)\rg|$, 
and the more familiar positivity conditions shed little light on the
size of the $u$-quark transversity distribution, simply that $\lf|\D_T u(x)\rg| \leqslant u(x)$.
However, the differing signs of the spin-averaged and helicity $d$-quark distributions
implies that for all values of $x$ where $\D d(x) \leqslant 0$ the transversity $d$-quark 
distribution must satisfy $\lf|\D_T d(x)\rg| \leqslant \tfrac{1}{2}d(x)$.
For the $d$-quark moments the Soffer inequality implies $\lf|\D_T d\rg| \leqslant \tfrac{1}{2}$,
provided $\D d \leqslant 0$. In fact, if $\D d \leqslant -\tfrac{1}{3}$,
which appears true empirically \cite{Gluck:2000dy,Hirai:2003pm,Bluemlein:2002be},
then the transversity moment must satisfy $\lf|\D_T d\rg| \leqslant \tfrac{1}{3}$.
Our results, presented in Figs.~\ref{fig:freeup} and \ref{fig:freedown}, 
satisfy all positivity constraints, including the Soffer bound.

\begin{figure}[tbp]
\centering\includegraphics[width=\columnwidth,clip=true,angle=0]{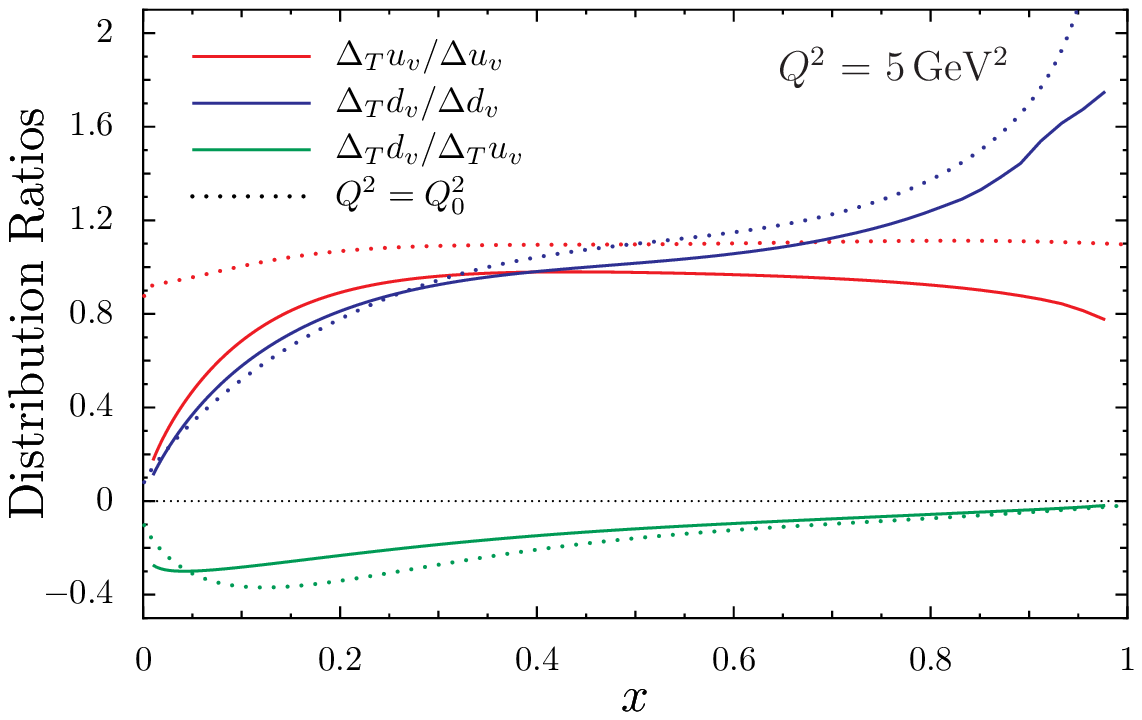}
\caption{Quark distribution ratios, in each case the dotted line is the result 
before QCD evolution and the solid line is the result for $Q^2=5\,$GeV$^2$.}
\label{fig:ratio3}
\end{figure}

The relative importance of each diagram in Fig.~\ref{fig:feydiagrams} to the 
quark distributions is easily illustrated if we consider their contributions
to the transversity moments. For the $u$-quark distributions we have
$\D_T f^s_{u/N} = 0.75$, $\D_T f^a_{u/N} = -0.016$, $\D_T f^s_{u(D)/N} = 0$, $\D_T f^a_{u(D)/N} = 0.090$
and $\D_T f^m_{u(D)/N} = 0.22$, where the isospin factors in Eq.~\eqref{eq:delu}
have been incorporated into the numerical values.
Results for the $d$-quark are obtained by dividing out the $u$-quark
isospin factors and multiplying by the $d$-quark factors from Eq.~\eqref{eq:deld}. 
The dominant terms for
$\D_T u(x)$ are clearly the scalar-quark and diquark-mixing diagrams. The strength
of $d$-quark distribution arises almost exclusively from the diquark-mixing term.
Similar results are also obtained for the helicity distributions \cite{Cloet:2005pp}.
This clearly illustrates the importance of including axial-vector diquark correlations 
in any quark-diquark picture of the nucleon.

In Figs.~\ref{fig:ratio2} and \ref{fig:ratio3} we illustrate various ratios of the 
transversity distributions with the unpolarized and helicity distributions.
In each case the dotted line is the result at the model scale, $Q^2=Q_0^2$, and the solid 
line is the ratio evolved to $Q^2 = 5\,$GeV$^2$. In Fig.~\ref{fig:ratio2} we present
our results for the single flavour ratios of transversity to unpolarized quark distributions. 
Both the up and down quark ratios are finite in the $x \to 1$ limit at the model scale. However,
as all transversity moments decrease with increasing $Q^2$, we 
expect the ratios to decrease after QCD evolution, which is evident for the $Q^2=5\,$GeV$^2$
results.\footnote{The non-conservation of the nucleon tensor charge results from
a non-zero anomalous dimension in Mellin space, $\D_T \g_{qq}(n)$, at leading order.
The sign of the transversity anomalous dimension at both the leading and 
next-to-leading order is negative, and therefore the tensor charge drops with increasing $Q^2$. 
Moreover, since $\D_T \g_{qq}(n) < \D \g_{qq}(n)$ for all moments $n$, the transversity moments
decrease more rapidly than the helicity moments. 
This has important consequences for the QCD evolution of the transversity distributions.}

Ratios of transversity to helicity distributions are presented in Fig.~\ref{fig:ratio3}.
In the large $x$ limit we find that the $u$-quark ratio approaches
1.1 and the $d$-quark ratio becomes quite large.
The $d$-quark result is surprising. It may be an inherent feature of the model or 
possibly a consequence of the static approximation. The quark exchange diagram, absent in this
approximation, contributes dominantly in the scalar diquark channel, where 
a $d$-quark must be exchanged. Therefore, including this contribution may significantly alter 
the ratio $\D_T d(x)/\D d(x)$ at large $x$. 
This possibility illustrates the importance of going beyond the static approximation in future work.
The final result illustrated in Fig.~\ref{fig:ratio3} is the mixed flavour transversity 
ratio, $\D_T d_v(x)/\D_T u_v(x)$, which is small for all $x$ and tends to zero as $x$ approaches one.

To aid comparison between our calculations and future investigations 
of the transversity distributions, we provide a parametrization of our
$u$- and $d$-quark valence distributions at the GRV scale, $Q^2 = 0.4\,$GeV$^2$. The
general form of the parametrization is
\begin{equation}
x\,\D_T q_v(x) = \eta x^\a(1-x)^\beta(1+\g\,x^\delta + \rho x).
\end{equation}
For the $u$-quark distribution we find
\begin{multline}
\eta   = 1.5025, \quad
\alpha = 1.1586, \quad
\beta  = 3.9940, \\
\gamma = 113.65, \quad
\delta = 11.150, \quad
\rho   = 12.214,
\end{multline}
and a fit to the $d$-quark distribution gives
\begin{multline}
\eta   = -2.7990, \quad
\alpha =  1.5952, \quad
\beta  =  5.7531, \\
\gamma = 45.424,  \quad
\delta =  8.0163, \quad
\rho   =  2.2065.
\end{multline}
These numbers produce an extremely good fit to our results, however one
should not read too much into their values as a similar $\chi^2$ can be achieved with
other sets of parameters. The transversity parametrization used in
Ref.~\cite{Anselmino:2007fs} has the form
\begin{multline}
x\,\D_T q_v(x) = \frac{1}{2}N^T_q\,x^\a(1-x)^\beta\\
\frac{\lf(\a+\beta\rg)^{\lf(\a+\beta\rg)}}{\a^\a\beta^\beta}\lf[q(x)+\D q(x)\rg],
\end{multline}
where $q(x)$ and $\D q(x)$ are the GRV parametrizations given in 
Refs.~\cite{Gluck:1998xa,Gluck:2000dy}. Using this parametrization, a fit to
our distributions at $Q^2 = 0.4\,$GeV$^2$ gives
\begin{equation}
N^T_u = 1.13, \quad N^T_d = -1.46, \quad  \a = 1.48, \quad \beta = 0.70,
\end{equation}
which also furnishes a good representation of our results. These values
are very different from those obtained in Ref.~\cite{Anselmino:2007fs}, reflecting
the fact that our distributions are considerably larger.

\section{Conclusion}

Using a covariant quark-diquark model for the nucleon,
which includes both the scalar and axial-vector diquark channels,
we calculated the transversity quark distributions for the proton.
A key feature of this framework is that it produces quark distributions 
that have the correct support and obey the number and momentum sum rules.
This model also eliminates unphysical thresholds for nucleon decay into quarks
and hence incorporates important aspects of confinement.

We found that at the model scale the transversity and helicity distributions
are very similar in magnitude, with $\D_T u_v(x)$ being slightly larger than
$\D u_v(x)$ for all $x$, whereas $\D_T d_v(x)$ becomes smaller than $\D d_v(x)$
for $x \lesssim 0.35$. These results satisfy the Soffer inequality.
For the first moments of the distributions we find
$\D_T u_v = 1.04$ and $\D_T d_v = -0.24$, giving a nucleon isovector tensor charge of $g_T = 1.28$
at the model scale. This result is very near the empirical value for the axial charge, 
$g_A = 1.267$, which we use as a constraint to help determine the model parameters.
Axial-vector diquarks were found to play a pivotal role, essentially through
the scalar--axial-vector mixing term included here for the first time in a
transversity calculation. 
This diagram almost exclusively gives rise to
the $d$-quark distribution and provides about ~20\% of the strength for the $u$-quark
transversity result. Future work in this direction will be to calculate fragmentation 
functions, in particular the Collins functions, so that a direct comparison between 
semi-inclusive DIS experiments and theory is possible.

We compared our transversity results with the recent experimental extraction of
Anselmino \textit{et al.}, and in each case found that our
distributions lie slightly outside the one-sigma confidence interval
of their empirical parametrizations, in the valence quark region.
The potentially small magnitude of the empirical $\D_T u(x)$ distribution,
if confirmed by future experimental data, would indicate very large relativistic
effects for the quarks in the nucleon. This would be very difficult to
explain within relativistic models utilizing the concept of a constituent
quark mass.

\section*{Acknowledgments} 
\vspace{-0.2em}
IC thanks Jefferson Lab for its hospitality and support, as this
is where much of this research was completed.
This work was supported by: Department of Energy, Office of Nuclear Physics, 
contract no. DE-AC02-06CH11357,
under which UChicago Argonne, LLC, operates Argonne National Laboratory;
contract no. DE-AC05-84ER40150, under which JSA 
operates Jefferson Lab; and by the Grant in Aid for Scientific Research of the 
Japanese Ministry of Education, Culture, Sports, Science and Technology, 
Project no. C-19540306.


\end{document}